\documentclass[traditabstract]{aa}

\usepackage{txfonts}
\usepackage{graphicx}
\usepackage{natbib}
\usepackage{xspace}
\usepackage{url}
\bibpunct{(}{)}{;}{a}{}{,}

\begin{document}
\title{COSMOGRAIL: the COSmological MOnitoring of \\GRAvItational Lenses\thanks{Based on observations made with the 2.0-m Himalayan Chandra Telescope (Hanle, India), the 1.5-m AZT-22 telescope (Maidanak Observatory, Uzbekistan), and the 1.2-m Mercator Telescope. Mercator is operated on the island of La Palma by the Flemish Community, at the Spanish Observatorio del Roque de los Muchachos of the Instituto de Astrof\'isica de Canarias.}\fnmsep\thanks{Light curves are available at the CDS via anonymous ftp to cdsarc.u-strasbg.fr (130.79.128.5) or via http://cdsarc.u-strasbg.fr/viz-bin/qcat?J/A+A/557/A44, and on http://www.cosmograil.org.}} 
\subtitle{XIV. Time delay of the doubly lensed quasar SDSS~J1001$+$5027}
\titlerunning{COSMOGRAIL XIV -- Time delay of SDSS~J1001$+$5027}

\author{S.~Rathna Kumar\inst{\ref{iia}}
\and M.~Tewes\inst{\ref{epfl}}
\and C.~S.~Stalin\inst{\ref{iia}}
\and F.~Courbin\inst{\ref{epfl}}
\and I.~Asfandiyarov\inst{\ref{uz}}
\and G.~Meylan\inst{\ref{epfl}}
\and E.~Eulaers\inst{\ref{liege}}
\and T.~P.~Prabhu\inst{\ref{iia}}
\and P.~Magain\inst{\ref{liege}}
\and H.~Van Winckel\inst{\ref{leuven}}
\and Sh.~Ehgamberdiev\inst{\ref{uz}}
}
\authorrunning{Rathna Kumar et al.}

\institute{
Indian Institute of Astrophysics, II Block, Koramangala, Bangalore 560 034, India, \email{rathna@iiap.res.in} \label{iia}
\and
Laboratoire d'astrophysique, Ecole Polytechnique F\'ed\'erale de Lausanne (EPFL), Observatoire de Sauverny, 1290 Versoix, Switzerland\label{epfl}
\and
Ulugh Beg Astronomical Institute, Uzbek Academy of Sciences, Astronomicheskaya 33, Tashkent, 100052, Uzbekistan\label{uz}
\and
Institut d'Astrophysique et de G\' eophysique, Universit\' e de Li\`ege, All\'ee du 6 Ao\^ut, 17, 4000 Sart Tilman, Li\`ege 1, Belgium \label{liege}
\and
Instituut voor Sterrenkunde, Katholieke Universiteit Leuven, Celestijnenlaan 200B, 3001 Heverlee, Belgium  \label{leuven}
}

\date{Received 21 June 2013 / Accepted 8 July 2013}
\abstract{
This paper presents optical R-band light curves and the time delay of the doubly imaged gravitationally lensed quasar SDSS~J1001$+$5027 at a redshift of 1.838.
We have observed this target for more than six years, between March 2005 and July 2011, using the 1.2-m Mercator Telescope, the 1.5-m telescope of the Maidanak Observatory, and the 2-m Himalayan Chandra Telescope.  
Our resulting light curves are composed of 443 independent epochs, and show strong intrinsic quasar variability, with an amplitude of the order of 0.2 magnitudes. From this data, we measure the time delay using five different methods, all relying on distinct approaches. One of these techniques is a new development presented in this paper. All our time-delay measurements are perfectly compatible. By combining them, we conclude that image A is leading B by $119.3 \pm 3.3$ days ($1\sigma$, 2.8\% uncertainty), including systematic errors. It has been shown recently that such accurate time-delay measurements offer a highly complementary probe of dark energy and spatial curvature, as they independently constrain the Hubble constant. The next mandatory step towards using SDSS~J1001$+$5027 in this context will be the measurement of the velocity dispersion of the lensing galaxy, in combination with deep Hubble Space Telescope imaging.
}


\maketitle

\section{Introduction}

In the current cosmological paradigm, only a handful of para\-meters seem necessary to describe the Universe on the largest scales and its evolution over time.
Testing this cosmological model requires a range of experiments, characterized by 
different sensitivities to these parameters. These experiments, or cosmological probes, 
are all affected by statistical and systematic errors and none of them on its own can uniquely constrain the cosmological models. This is due to the degeneracies
inherent in each specific probe, implying that the probes become truly effective in constraining
cosmology only when used in combination.

The latest cosmology results by the Planck consortium beautifully illustrate this \citep{Planck_cosmo}.
In particular, the constraints obtained by Planck on the Hubble parameter $H_0$, on the curvature 
$\Omega_k$, and on the dark energy equation of state parameter $w$ rely mostly on the
combination of the baryonic acoustic oscillations measurements (BAO) with the 
Cosmic microwave background (CMB) observations.

Strong gravitational lensing offers a valuable yet inexpensive complement to independently constrain some of the cosmological parameters, through the measurement of the so-called time delays in quasars strongly lensed by a foreground galaxy \citep{refsdal1964}.
The principle of the method is the following. The travel times of photons along the distinct optical paths forming the multiple images are not identical. These travel-time differences, called the time delays, depend on the geometrical differences between the optical paths, which contain the cosmological information, and on the potential well of the lensing galaxy(ies). In practice, time delays can be measured from photometric light curves of the multiple images of lensed quasar: if the quasar shows photometric variations, these are seen in the individual light curves at epochs separated by the time delay.

A precise and accurate measurement of such a time delay, in combination with a well-constrained model for the lensing galaxy, can be used to  constrain cosmology in a way which is very complementary to other cosmological probes \citep[see, e.g., ][]{Linder:2011cs}. A recent and remarkable implementation of this approach can be found in \citet{Suyu2013} that uses the time-delay measurements from \citet{1131}. We note, however, that to obtain a robust cosmological inference from this time-delay technique, particular attention must be paid to any possible lens model degeneracies \citep{SS13, replytoSS13, SS13b}.

So far, only a few quasar time delays have been measured convincingly, from long and well-sampled light curves. 
The international COSMOGRAIL\footnote{\url{http://www.cosmograil.org/}} 
(COSmological MOnitoring of GRAvItational Lenses)
collaboration is changing this situation by measuring accurate time delays for a large
number of gravitationally lensed quasars. The goal of COSMOGRAIL is to reach an accuracy of less
than 3\%, including systematics, for most of its targets.

In this paper, we present the time-delay measurement 
for the two-image gravitationally lensed quasar  SDSS~J1001$+$5027 
($\alpha_{2000}$ = 10:01:28.61, $\delta_{2000}$ = +50:27:56.90) at $z=1.838$ \citep{2005ApJ...622..106O}. The image separation of $\Delta \theta = 2.86$\arcsec \citep{2005ApJ...622..106O} and the high declination 
of the target make it a relatively easy prey for medium-sized northern telescopes and average seeing conditions. 
The redshift of the lensing galaxy $z_l=0.415$ has been measured spectroscopically 
\citep{Inada2012}.
 
Our paper is structured as follows. Section \ref{section:monitoring} describes our monitoring, the data reduction, and the resulting light curves. In Sect. \ref{section:newtechnique} we present a new time-delay point estimator. We add this technique to a pool of four other existing algorithms, to measure the time delay in Sect. \ref{section:results}. Finally, we summarize our results and conclude in Sect. \ref{section:conclusion}.

\section{Observations, data reduction, and light curves}
\label{section:monitoring}

\begin{table*}[t!]
\caption{Summary of COSMOGRAIL observations of SDSS~J1001$+$5027.}
\label{table:observations}
\begin{center}
\begin{tabular}{l l c c c r r r}
\hline
\hline
Telescope & Camera & FoV & Pixel scale & Monitoring period & Epochs & Exp. time\tablefootmark{a} & Sampling\tablefootmark{b} \\
\hline
Mercator 1.2 m & MEROPE & $6.5\arcmin \times 6.5\arcmin$ & $0\farcs190$ & 2005 Mar -- 2008 Dec & 239 & 5 $\times$ 360 s &  3.8 (2.0) d \\
HCT 2.0 m & HFOSC & $10\arcmin \times 10\arcmin$ & $0\farcs296$ & 2005 Oct -- 2011 Jul & 143 & 4 $\times$ 300 s &  9.5 (6.1) d \\
Maidanak 1.5 m & SITE & $8.9\arcmin \times 3.5\arcmin$ & $0\farcs266$ & 2005 Dec -- 2008 Jul & 41 & 7  $\times$  180 s &  5.9 (4.1) d \\
Maidanak 1.5 m & SI & $18.1\arcmin \times 18.1\arcmin$ & $0\farcs266$ & 2006 Nov -- 2008 Oct & 20 & 6 $\times$ 600 s & 12.6 (9.5) d \\
\hline
Combined &  &  &  & 2005 Mar -- 2011 Jul &  443 & 201.5 h &  3.8 (1.9) d \\
\hline
\end{tabular}
\end{center}
\tablefoot{
\tablefoottext{a}{The exposure time is given by the number of dithered exposures per epoch and their individual exposure times.}
\tablefoottext{b}{The sampling is given as the mean (median) number of days between two consecutive epochs, excluding the seasonal gaps.}
}
\end{table*}

\subsection{Observations}

\begin{figure}[tbp]
\resizebox{\hsize}{!}{\includegraphics{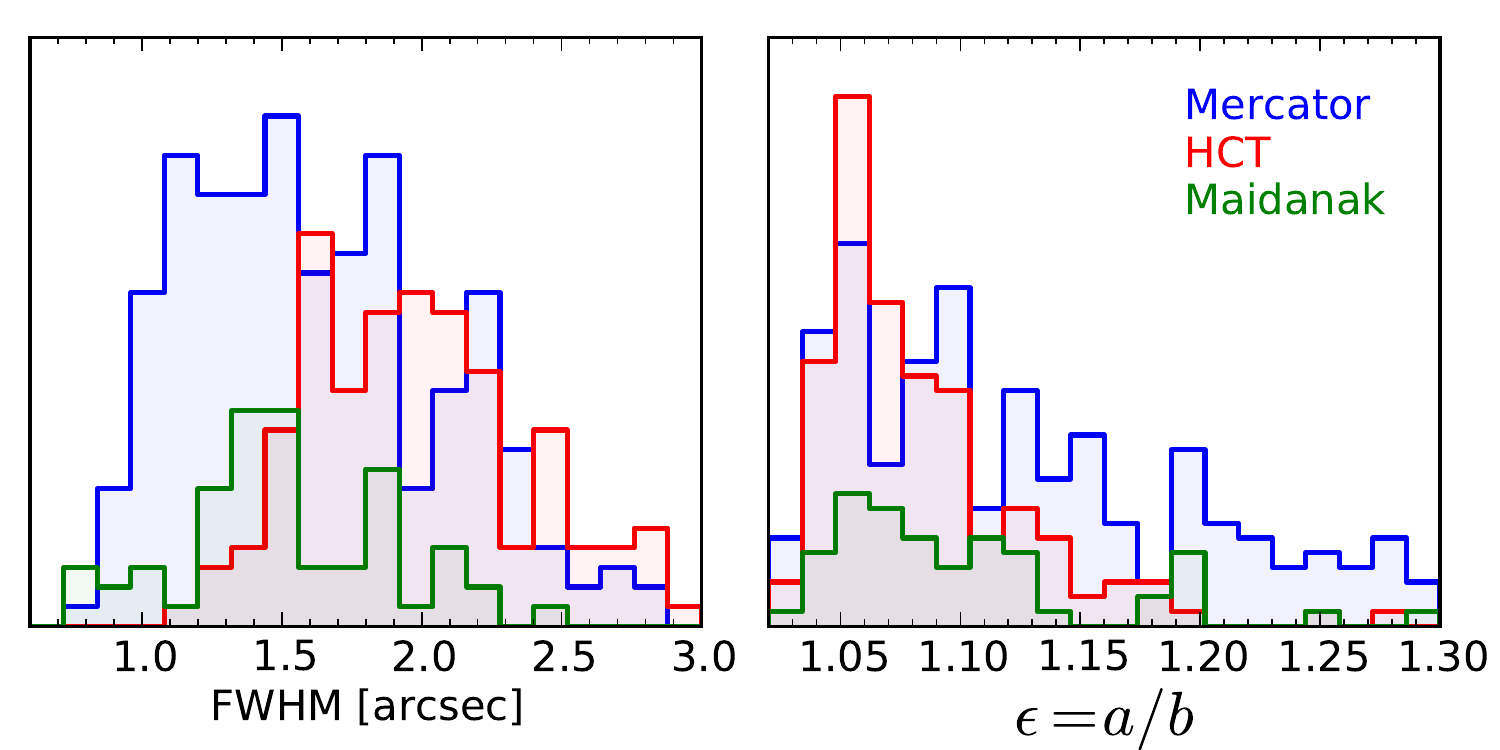}}
\caption{Distribution of the average observed FWHM and elongation $\epsilon$ of field stars in the images
used to build the light curves of SDSS~J1001$+$5027.}
\label{figure:fwhm}
\end{figure}

We monitored SDSS~J1001$+$5027 in the R band for more than six years, from March 2005 to July 2011, with three different telescopes: the 1.2-m Mercator Telescope located at the Roque de los Muchachos Observatory on La Palma (Spain), the 1.5-m telescope of the Maidanak Observatory in Pamir Alai (Uzbekistan), and the 2-m Himalayan Chandra Telescope (HCT) located at the Indian Astronomical Observatory in Hanle (India). Table \ref{table:observations} details our monitoring observations. In total we obtained photometric measurements for 443 independent epochs, with a mean sampling interval below four days. Each epoch consists of at least three, but mostly four or more, dithered exposures. Figure \ref{figure:fwhm} summarizes the image quality of our data. The COSMOGRAIL collaboration has now ceased the monitoring of this target, to focus on other systems.

\subsection{Deconvolution photometry}
\label{section:deconvolution}

\begin{figure*}[htbp]
\begin{center}
\resizebox{\hsize}{!}{\includegraphics{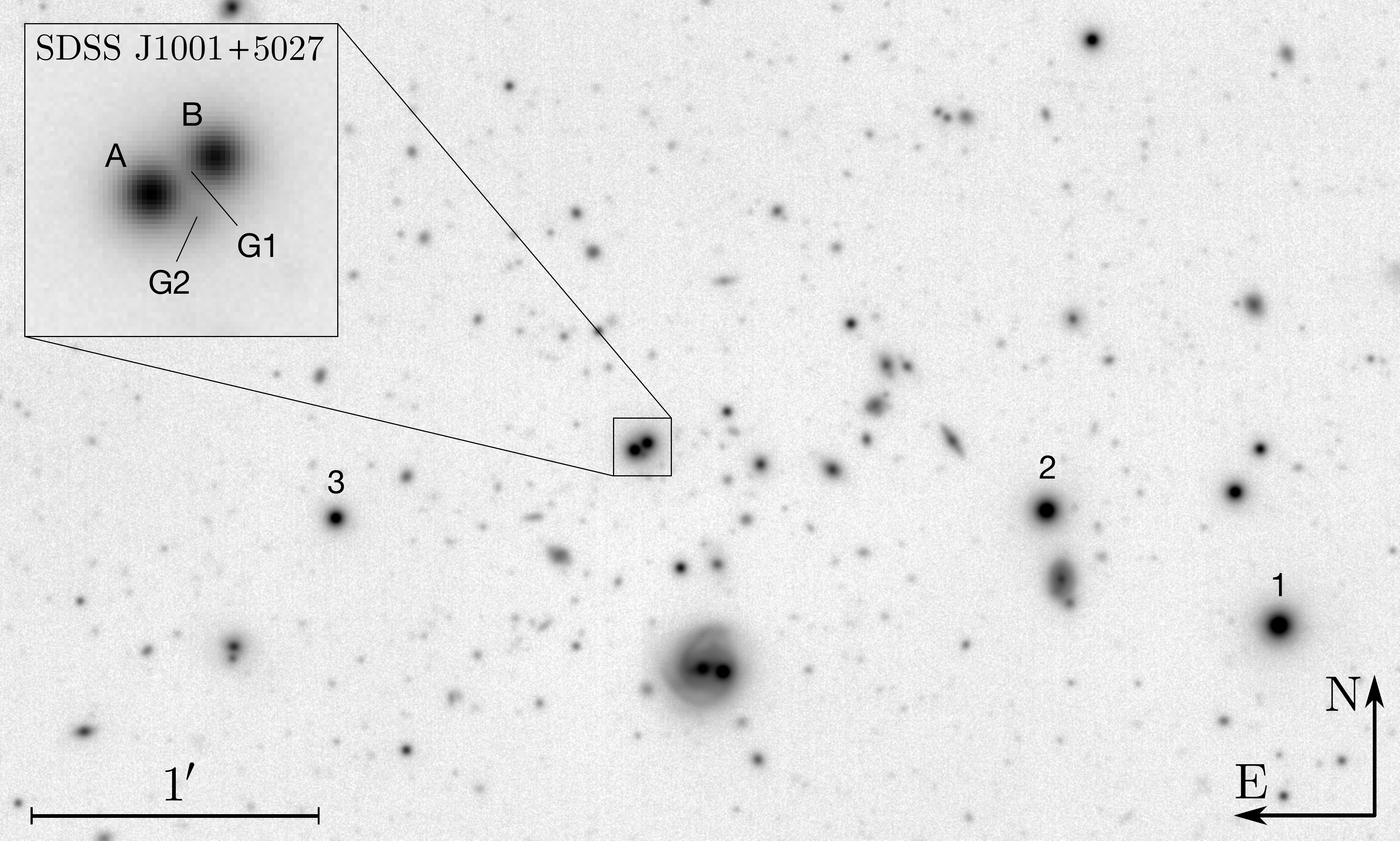}}
\caption{R-band image centered on SDSS~J1001$+$5027. The image is the combination of the 210 best exposures from the Mercator telescope, for a total exposure time of 21 hours. We use the stars labeled 1, 2, and 3 to model the PSF and to cross-calibrate the photometry of each exposure. The position of the two lensing galaxies G1 and G2 are indicated in the zoomed image in the upper left. They are most clearly seen in the deconvolved images presented in Fig.~\ref{figure:dec}.}
\label{figure:field}
\end{center}
\end{figure*}

\begin{figure*}[htbp]
\begin{center}
\resizebox{0.8 \hsize}{!}{\includegraphics{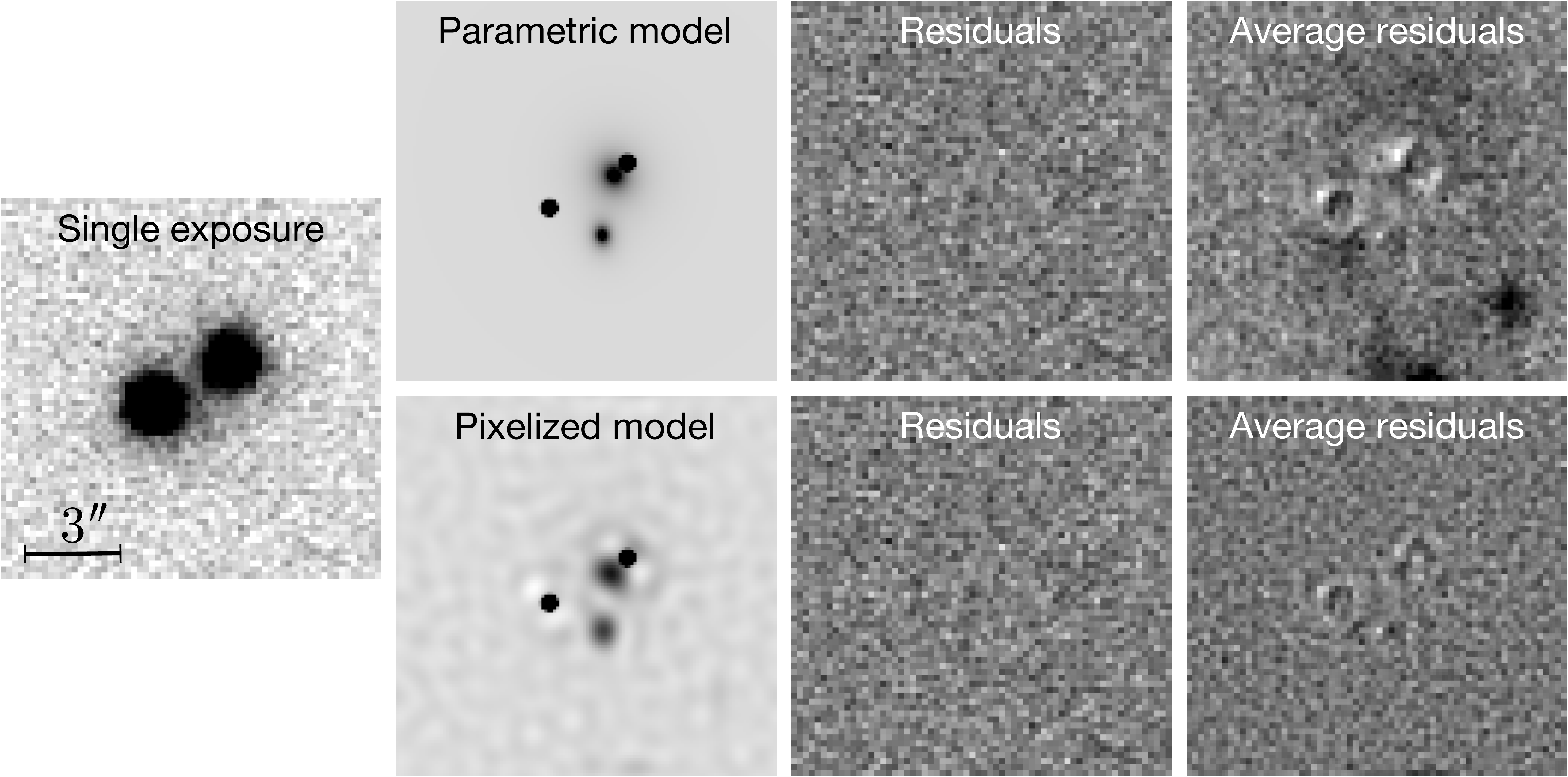}}
\caption{Two ways of modeling the light distribution for extended objects during the deconvolution process. On the left is shown a single 360-second exposure of SDSS~J1001$+$5027 obtained with the Mercator telescope in typical atmospheric conditions. The other panels show the parametric (top row) and pixelized light models (bottom row) for the lens galaxies as described in the text. The residual image for the single exposure is also shown in each case, as well as the average residuals over the 120 best exposures. The residual maps are normalized by the shot noise amplitude. The dark areas indicate excess flux in the data with respect to the model. Gray scales are linear.
}
\label{figure:dec}
\end{center}
\end{figure*}

The image reduction and photometry closely follows the procedure described in \citet{1131}. We performed the flat-field correction and bias subtraction for each exposure using custom software pipelines, which address the particularities of the different telescopes and instruments. 

Figure \ref{figure:field} shows part of the field around SDSS~J1001$+$5027, obtained by stacking the best monitoring exposures from the Mercator telescope to reach an integrated exposure time of 21 hours. The relative flux measurements of the quasar images and reference stars for each individual epoch were obtained through our COSMOGRAIL photometry pipeline, which is based on the simultaneous MCS deconvolution algorithm \citep{1998ApJ...494..472M}. The stars labeled 1, 2, and 3 in Fig. \ref{figure:field} are used to characterize the point spread function (PSF) and relative magnitude zero-point of each exposure.

The two quasar images A and B of SDSS~J1001$+$5027 are separated by $2.86\arcsec$, which is significantly larger than the typical separation in strongly lensed quasars. In principle, this makes SDSS~J1001$+$5027 a relatively easy target to monitor, as the quasar images are only slightly blended in most of our images. However, image B lies close to the primary lensing galaxy G1. Minimizing the additive contamination by G1 to the flux measurements of B therefore requires a model for the light distribution of G1. In Fig. \ref{figure:dec}, we show two different ways of modeling these galaxies. Our standard approach, shown in the bottom panels, consists in representing all extended objects, such as the lens galaxies, by a regularized pixel grid. The values of these pixels get iteratively updated during the deconvolution photometry procedure. Because of obvious degeneracies, this approach may fail when a relatively small extended object (lens galaxy) is strongly blended with a bright point source (quasar), leading to unphysical light distributions. To explore the sensitivity of our results to a possible bias of this kind, we have adopted an alternative approach of representing G1 and G2 by two simply parametrized elliptical Sersic profiles, as shown in the top row of Fig. \ref{figure:dec}. For both cases, the residuals from single exposures are convincingly homogeneous. Only when averaging the residuals of many exposures to decrease the noise can the simply parametrized models be seen to yield a less good overall fit to the data, since they cannot represent additional background sources nor compensate for small systematic errors in the shape of the PSF. 

We find that the difference between these approaches in terms of the resulting quasar flux photometry is marginal; it is insignificant regarding the measurement of the time delay. In all the following we will use the quasar photometry obtained using the parametrized model (top row of Fig. \ref{figure:dec}) which is likely to be closer to reality than our pixelized model in the immediate surroundings of image B.

\subsection{Light curves}
\label{section:lightcurves}

\begin{figure*}[htbp]
\begin{center}
\resizebox{\hsize}{!}{\includegraphics{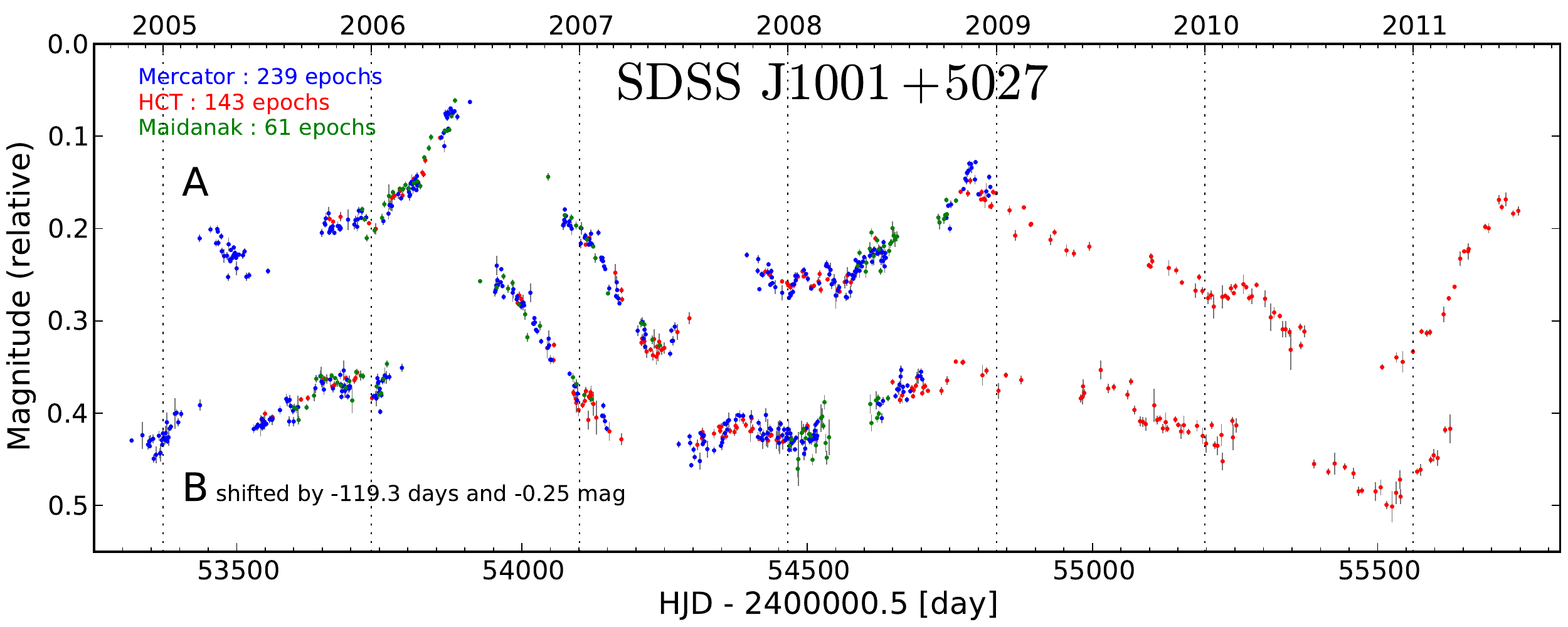}}
\caption{R-band light curves of the quasars images A and B in SDSS~J1001$+$5027 from March 2005 to July 2011. The 1$\sigma$ photometric error bars are also shown. For display purpose, the curve of quasar image B is shown shifted in time by the measured time delay (see text). The light curves are available in tabular form from the CDS and the COSMOGRAIL website.
}
\label{figure:lightcurves}
\end{center}
\end{figure*}

Following \citet{1131}, we empirically corrected for small magnitude and flux shifts between the light curve contributions from different telescopes/cameras to obtain minimal dispersion in each of the combined light curves. In the present case we chose the photometry from the Mercator telescope as a reference, and for the data from the Maidanak and HCT telescopes, we optimized a common magnitude shift and individual flux shifts for A and B.

Figure \ref{figure:lightcurves} shows the combined 6.5-season long light curves, from which we measure a time delay of $\Delta t_{\mathrm{AB}} = -119.3$ days (see Sect.~\ref{section:results}). In this figure, light curve B has been shifted by this time delay to highlight the correspondence and temporal overlap of the data.
We observe strong intrinsic quasar variability, common to images A and B. In the period 2006 to 2007, the variability in image A is as large as 0.25 magnitudes over a single year. In addition to this large scale correspondence, several small and short scale intrinsic variability features are common to both curves, for instance around December 2005 and January 2010. Our data unambiguously reveal, already to the eye, an approximate time delay of $\Delta t_{\mathrm{AB}} \approx -120$ days, with A leading B.

\subsection{An apparent mismatch between the light curves of the quasar images}
\label{mismatch}

The apparent flux ratio between the quasar images, as inferred from the time-shifted light curves shown in Fig. \ref{figure:lightcurves}, stays roughly in the range from 0.40 to 0.44 mag over the length of our monitoring. Strong gravitational lens models readily explain different magnifications of the quasar images, yielding stationary flux ratios or magnitude shifts between the light curves. Figure \ref{figure:lightcurves} hints, however, at a moderate correlation between some variable flux ratio and the intrinsic quasar variability. In particular, the amplitude of the quasar variability, in units of magnitudes, appears to be smaller in B than in A. Potential reasons for this mismatch include the effects of microlensing by stars of the lens galaxy, or a contamination of the photometry of B by some additive external flux. We find that one has to subtract from curve B about 20\% of its median flux to obtain an almost stationary magnitude shift of about 0.66 mag between the light curves. As this contamination would be several times larger than the entire flux of galaxy G1, we conclude that plausible errors of our light models for G1 cannot be responsible for the observed discrepancy between the light curves.

\section{A new time-delay estimator}
\label{section:newtechnique}

Although an unambiguous approximation of the time delay of SDSS~J1001$+$5027 can be made by eye, accurately measuring its value is not trivial, and is made more difficult by the extrinsic variability between the light curves. Even more obvious features of the data, such as the sampling gaps due to non-visibility periods of the targets, could easily bias the results from a time-delay measurement technique. The impact of these effects on the quality of the time-delay inference clearly differs for each individual quasar lensing system and dataset. To check for potential systematic errors, we feel that a wise approach is to employ several numerical methods based on different fundamental principles.

In the present section we introduce a new time-delay estimation method, based on minimizing residuals of a high-pass filtered difference light curve between the quasar images.

\subsection{The difference-smoothing technique}
\label{section:diff-smooth}

\begin{figure*}[t!]
\resizebox{\hsize}{!}{\includegraphics{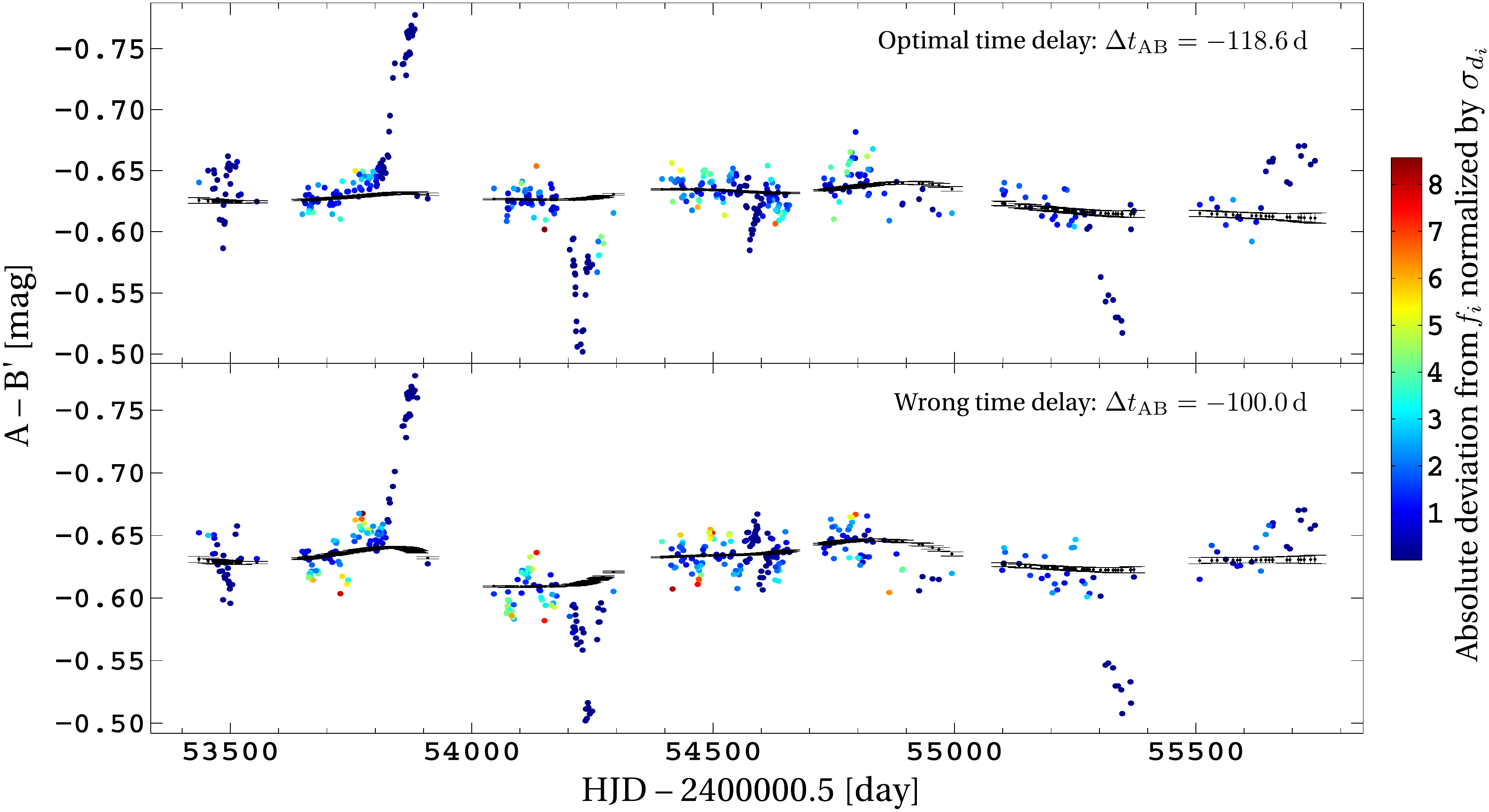}}
\caption{Difference light curves of SDSS~J1001$+$5027 as obtained by the new difference-smoothing technique introduced in this paper. The curves are shown for the best time-delay estimate found with this technique (top panel, $\Delta t_{\mathrm{AB}} = - 118.6$ days), and for a wrong time-delay value (bottom panel, $\Delta t_{\mathrm{AB}} = - 100.0$ days). The difference light curves $d_i$ are shown as colored points. They are smoothed using a kernel of width $s=100$ days to compute the $f_i$ (black points). The error bars on the black points show the uncertainty coefficients $\sigma_{f_i}$. The points in the difference light curve $d_i$ are color-coded according to the absolute factors of their uncertainties $\sigma_{d_i}$ by which they deviate from $f_i$. In both panels, light curve A is used as reference, and light curve B is shifted in flux by the same amount.}
\label{figure:microlensing}
\end{figure*}

This technique is a point estimator that determines both an optimal time delay and an optimal shift in \emph{flux} between two light curves, while also allowing for smooth extrinsic variability. The correction for a flux shift between the light curves explicitly addresses the mismatch described in Sect. \ref{mismatch}, whatever its physical explanation. This flux shift may be due to a contamination of light curve B by residual light from the lensing galaxy, from the lensed quasar host galaxy, or by microlensing resolving the quasar structure.

We consider two light curves A and B sampled at epochs $t_i$, where A$_i$ and B$_i$ are the observed magnitudes at epochs $t_i$, ($i=1,2,3,...,N$). We select A as the reference curve. Light curve B is shifted in time with respect to A by some amount $\tau$, and in \emph{flux} by some amount $\Delta f$.  Formally, this shifted version B$^\prime$ of B is given by
\begin{eqnarray}
\mathrm{B}_i^\prime&=&-2.5\, {\rm log} \left( 10^{-0.4\, \mathrm{B}_i} + \Delta f \right), \\
t_i^\prime&=&t_i + \tau.
\end{eqnarray}
For any estimate of the time delay $\tau$ and of the flux shift $\Delta f$, we form a \emph{difference light curve}, with points $d_i$ at epochs $t_i$,
\begin{equation}
\label{equation:diffcurve}
d_i(\tau, \Delta t)= \mathrm{A}_i- \frac{\sum_{j=1}^N w_{ij} \mathrm{B}_j^\prime}{\sum_{j=1}^N w_{ij}} , 
\end{equation}
where the weights $w_{ij}$ are given by
\begin{equation}
\label{equation:pairing}
w_{ij} = \frac{1}{\sigma_{\mathrm{B}_j}^2} e^{-(t_j^\prime-t_i)^2 / 2\delta^2}. 
\end{equation}
The parameter $\delta$ is the decorrelation length, as in \citet{1996A&A...305...97P}, and $\sigma_{\mathrm{B}_j}$ denotes the photometric error of the magnitude B$_j$. This decorrelation length should typically be of the order of the sampling period, small enough to not smooth out any intrinsic quasar variability features from the light curve B. The uncertainties on each $d_i$ are then calculated as
\begin{equation}
\sigma_{d_i} = \sqrt{\sigma_{\mathrm{A}_i}^2+\frac{1}{\sum_{j=1}^N w_{ij}}},
\end{equation}
where $w_{ij}$ are given by Eq.~\ref{equation:pairing}. To summarize, at this point we have a discrete difference light curve, sampled at the epochs of curve A, built by subtracting from light curve A a smoothed and shifted version of B. We now smooth this difference curve $d_i$, again using a Gaussian kernel, to obtain a model $f_i$ for the differential extrinsic variability
\begin{equation}
f_i = \frac{\sum_{j=1}^N \nu_{ij} \, d_j}{\sum_{j=1}^N \nu_{ij}},
\end{equation} 
where the weights $\nu_{ij}$ are given by
\begin{equation}
\label{equation:smoothing}
\nu_{ij} = \frac{1}{\sigma_{d_j}^2}{e^{-(t_j-t_i)^2 / 2s^2}}.
\end{equation}
The smoothing time scale $s$ is a second free parameter of this method. Its value must be chosen to be significantly larger than $\delta$. 
For each $f_i$, we compute an uncertainty coefficient
\begin{equation}
\sigma_{f_i} = \sqrt{\frac{1}{\sum_{j=1}^N \nu_{ij}}}.
\end{equation} 
The idea of the present method is now to optimize the time-delay estimate $\tau$ and flux shift $\Delta f$ to minimize residuals between the difference curve $d_i$ and the much smoother $f_i$. Any incorrect value for $\tau$ introduces relatively fast structures that originate from the quasar variability into $d_i$, and these structures will not be well represented by $f_i$.
Figure \ref{figure:microlensing} illustrates this phenomenon in the case of SDSS~J1001$+$5027 by showing $d_i$ and $f_i$ for an optimal and an arbitrarily chosen wrong time-delay estimate. In both panels of Fig.~\ref{figure:microlensing}, the largest deviations between $d_i$ and $f_i$ are due to poorly constrained points with very high $\sigma_{d_i}$, and are therefore not significant. However, for the incorrect time-delay estimate, a larger number of well-constrained points of $d_i$ significantly deviate from $f_i$ (yellow and red points). To quantify this match between $d_i$ and $f_i$ we define a cost function in the form of a normalized $\chi^2$, 
\begin{equation}
\label{chi2}
\overline{\chi}^2 = \left[ \sum_{i=1}^N \frac{(d_i-f_i)^2}{\sigma_{d_i}^2+\sigma_{f_i}^2} \right] / \left[ \sum_{i=1}^N \frac{1}{\sigma_{d_i}^2+\sigma_{f_i}^2} \right],  
\end{equation}
and minimize this $\overline{\chi}^2(\tau, \Delta f)$ using a global optimization.

In the above description, light curves A and B are not interchangeable, thus introducing an asymmetry into the time-delay measurement process. To avoid this arbitrary choice of the reference curve, we systematically perform all computations for both permutations of A and B, and minimize the sum of the two resulting values of $\overline{\chi}^2$.

\subsection{The uncertainty estimation procedure}

As a point estimator, the technique described above does not provide information on the uncertainty of its result. We stress that simple statistical techniques such as variants of bootstrapping or resampling cannot be used to quantify the uncertainty of such highly non-linear time-delay estimators \citep{pycs}. These approaches are not able to discredit ``lethargic'' estimators, which favor a particular solution (or a small set of solutions) while being relatively insensitive to the actual shape of the light curves. Furthermore, they are not sensitive to plain systematic biases of the techniques.

Consequently, to quantify the random and systematic errors of this estimator, for each dataset to be analyzed and as a function of its free parameters, we follow the Monte Carlo analysis described in \citet{pycs}. It consists in applying the point estimator to a large number of fully synthetic light curves, which closely mimic the properties of the observed data, but have known true time delays. It is particularly important that these synthetic curves cover a range of true time delays around a plausible solution, instead of all having the same true time delay. Only this feature enables the method to adequately penalize estimators with lethargic tendencies.

\subsection{Application to SDSS~J1001$+$5027}
\label{application}

The decorrelation length $\delta$ and the width of the smoothing kernel $s$ are the two free parameters of the described technique. In this work, we choose $\delta$ to be equal to the mean sampling of the light curves ($\delta = 5.2$ days) and $s = 100$ days, yielding a point estimate of $\Delta t_{\mathrm{AB}} = - 118.6$ days for the time delay. The corresponding $d_i$ and $f_i$ difference light curves are shown in the top panel of Fig.~\ref{figure:microlensing}. Results of the uncertainty analysis will be presented in the next section, together with the performance of other point estimators.

We have explored a range of alternative values for the free parameters ($s = 50, 100, 150, 200$ and $\delta = 2.6, 5.2, 10.4$ days), and find that neither the time-delay point estimate from the observed data, nor the error analysis is significantly affected. The time-delay estimates resulting from these experiments stay within 1.2 days around the reference value obtained for $\delta = 5.2$ and $s=100$ days. Regarding the uncertainty analysis, we observe that increasing the smoothing length scale $s$ beyond 100 days decreases the random error, but at the cost of an increasing bias, which is not surprising.

\section{Time-delay measurement of SDSS~J1001$+$5027}
\label{section:results}

\begin{figure}[t]
\begin{center}
\resizebox{1.0 \hsize}{!}{\includegraphics{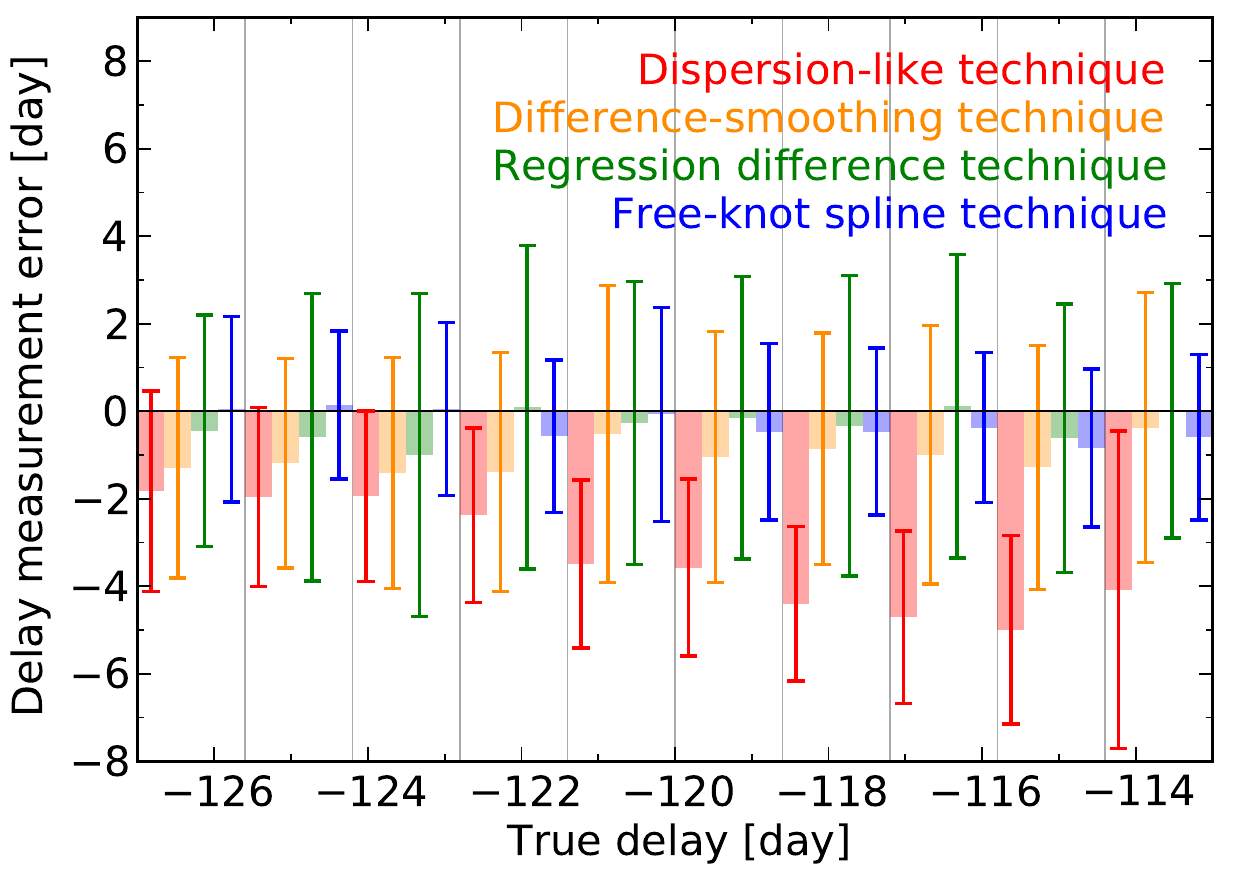}}
\caption{Error analysis of the four time-delay measurement techniques, based on delay estimations on 1000 synthetic curves that mimic our SDSS~J1001$+$5027 data. The horizontal axis corresponds to the value of the true time delay used in these synthetic light curves. The gray vertical lines delimit bins of true time delay. In each of these bins, the colored rods and 1$\sigma$ error bars show the systematic biases and random errors, respectively, as committed by the different techniques.}
\label{figure:measvstrue}
\end{center}
\end{figure}

\begin{figure}[t]
\begin{center}
\resizebox{1.0 \hsize}{!}{\includegraphics{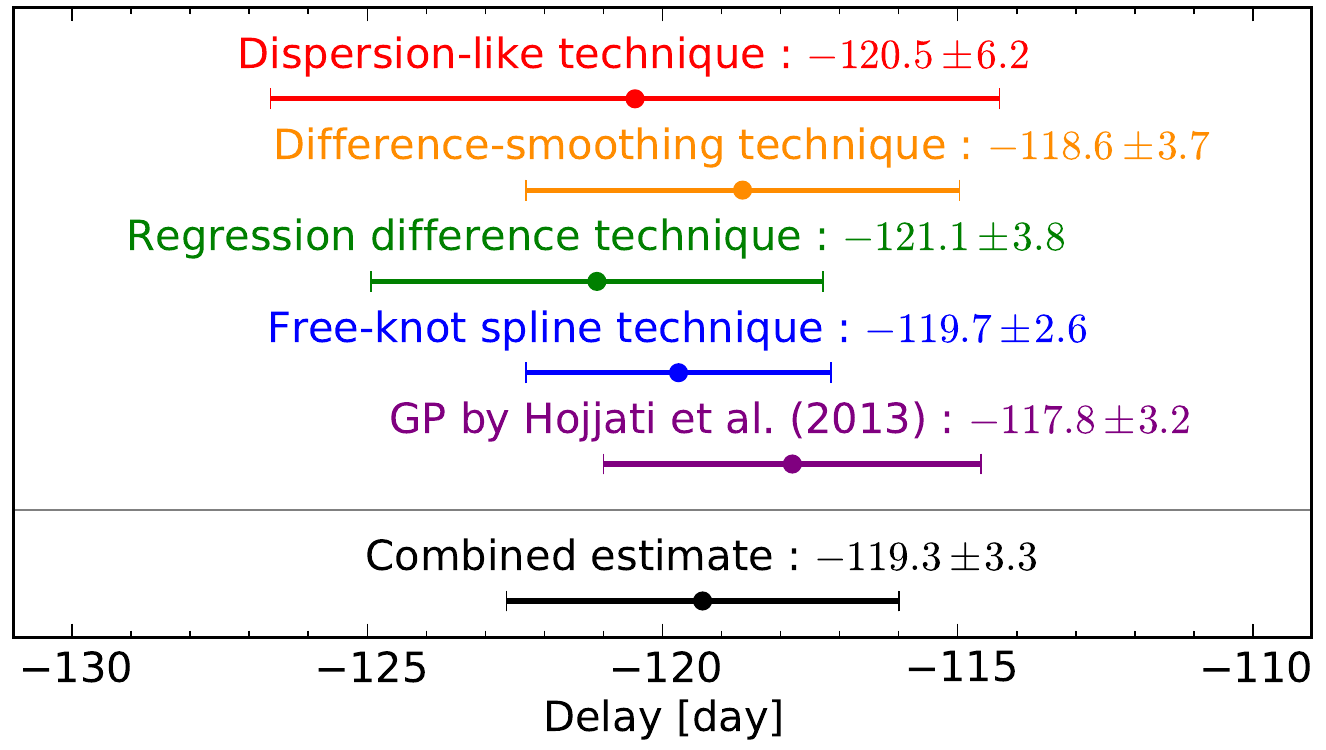}}
\caption{Time-delay measurements of SDSS~J1001$+$5027, following five different methods. The total error bar shown
here includes systematic and random errors.}
\label{figure:delays}
\end{center}
\end{figure}

\begin{table}
\caption{Time-delay measurements for SDSS~J1001$+$5027. The total $1\sigma$ error bars are given. Whenever possible,
we give in parenthesis the breakdown of the error budget: (random, systematic).}
\label{table:delays}
\centering
\begin{tabular}{l l}
\hline\hline
Method & $\Delta t_{\mathrm{AB}}$ [day]\\
\hline
Dispersion-like technique & -120.5 +/- 6.2 (3.6, 5.0)\\
Difference-smoothing technique & -118.6 +/- 3.7 (3.4, 1.4)\\
Regression difference technique & -121.1 +/- 3.8 (3.7, 1.0)\\
Free-knot spline technique & -119.7 +/- 2.6 (2.4, 0.8)\\
GP by Hojjati et al. (2013) & -117.8 +/- 3.2\\
\hline
Combined estimate (see text) & -119.3 +/- 3.3\\
\hline
\end{tabular}
\end{table}

In this work, we use five different methods to measure the time delay of SDSS~J1001$+$5027 from the data shown in Fig. \ref{figure:lightcurves}. All these methods have been developed to address light curves affected by extrinsic variability, resulting from microlensing or flux contamination. Three of the techniques, namely the dispersion-like technique, the regression difference technique, and the free-knot spline technique are described in length in \citet{pycs} and were used to measure the time delays in the four-image quasar RX J1131$-$123 \citep{1131}. 

In the the previous section, we presented our fourth method, the difference-smoothing technique. These first four methods are point estimators: they provide best estimates, without information on the uncertainty of their results. We proceed by quantifying the accuracy and precision of these estimators by applying them to a set of 1000 fully synthetic light curves, produced and adjusted following \citet{pycs}. These simulations include the intrinsic variations of the quasar source, mimicking the observed variability of SDSS~J1001$+$5027, as well as extrinsic variability on a range of time scales from a few days to several years. They share the same sampling and scatter properties as the real observations.

Figure \ref{figure:measvstrue} shows the results of this analysis, depicting the delay measurement error as a function of the true delay used to generate the synthetic light curves. As always, this analysis naturally takes into account the intrinsic variances of the techniques, that are due to the limited ability of the employed global optimizers to find the absolute minima of the cost functions.

As can be seen in Fig. \ref{figure:measvstrue}, the dispersion-like technique is strongly biased for this particular dataset. This could be a consequence of the simplistic polynomial correction for extrinsic variability linked to this technique. For the other techniques, the bias remains smaller than the random error, and no strong dependence on the true time delay is detected.

The final systematic error bar for each of these four techniques is taken as the worst measured systematic error on the simulated light curves (biggest colored rod in Fig.~\ref{figure:measvstrue}). The final random error is taken as the largest random error across the range of tested time delays. Finally, the total error bar for each technique is obtained by summing the systematic and random components in quadrature.

In the writing process of this paper, \cite{Hojjati2013} proposed a new independent method to measure time delays that is also able to address extrinsic variability. Their method is based on Gaussian process modeling, and does not rely on point estimation. It provides its own standalone estimate of the total uncertainty. We have provided these authors with the COSMOGRAIL data of SDSS~J1001$+$5027, without letting them know our measured values. They find $\Delta t_{\mathrm{AB}} = -117.8 \pm 3.2$ days.

We include this measurement by \cite{Hojjati2013} as a fifth measurement in our result summary, presented in Table~\ref{table:delays} and in a more graphical form in Fig.~\ref{figure:delays}. Not only do their time-delay values agree with our four estimates, but also their error bars agree well with ours, in spite of the totally different way of estimating them. 

We have five time-delay estimates from five very different methods, and all these estimates are compatible with each other. We now need to combine these results. In doing this, we exclude the delay from the dispersion-like technique that, as we show, is dominated by systematic errors. While the estimates from the four remaining techniques are obtained with very different methods, they are still not independent, as they all make use of the same data. We therefore simply average the four time-delay measurements to obtain our combined estimate, and we use the average of the total uncertainties as the corresponding uncertainty. This leads to $\Delta t_{\mathrm{AB}} = -119.3 \pm 3.3$ days, shown in black in Fig.~\ref{figure:delays}.

\section{Conclusion}
\label{section:conclusion}

In this paper, we present the full COSMOGRAIL light curves for the two images of the gravitationally lensed quasar SDSS~J1001$+$5027. The final data, all taken in the R band, totalize 443 observing epochs, with a mean temporal sampling of 3.8 days, from the end of 2004 to mid-2011. The COSMOGRAIL monitoring campaign for SDSS~J1001$+$5027 is no longer in progress. It involved three different telescopes with diameters from 1.2 m to 2 m, hence illustrating the effectiveness of small telescopes in conducting long-term projects with potentially high impact on cosmology. 

We analyzed our light curves with five different numerical techniques, including the three methods described in \citet{pycs}. In addition, we introduced and described a new additional method, based on representing the extrinsic variability by a smoothed version of the difference light curve between the quasar images. Finally, we also presented results obtained via the technique of \citet{Hojjati2013}, based on modeling of the quasar and microlensing variations using Gaussian processes. The  technique was \emph{blindly} applied to the data by the authors of \citet{Hojjati2013}, without any prior knowledge of the results obtained with the other four methods.

Aside from the dispersion-like technique, dominated by systematic errors, we find that the four other methods yield similar time-delay values and similar random and systematic error bars.  Our final estimate of the time delay is taken as the mean of these four best results, together with the mean of their uncertainties: $\Delta t_{\mathrm{AB}} = -119.3 \pm 3.3$ days, with image A leading image B. This is a relative uncertainty of 2.8\%, including systematic errors. 

The present time-delay measurement can be used in combination with lens models to constrain cosmological parameters, in particular the Hubble parameter H$_0$ and the curvature $\Omega_k$ \cite[e.g.,][]{Suyu2013}. 
The accuracy reached on cosmology with SDSS~J1001$+$5027 alone or in combination with other lenses, will rely on the availability of follow-up observations to measure: (1) the lens velocity dispersion, (2) the mass contribution of intervening objects along the line of sight, and (3) the detailed structure of the lensed host galaxy of the quasar. This translates in practice into one single night of an 8m-class telescope, plus about four orbits of the Hubble Space Telescope. 

\begin{acknowledgements}
We thank the numerous observers who contributed to the data from the Mercator and Maidanak telescopes, and we are grateful for the support provided by the staff at the Indian Astronomical Observatory, Hanle and CREST, Hoskote. We also thank A. Hojjati, A. Kim, and E. Linder for running their curve shifting algorithm on our data.  S. Rathna Kumar and C. S. Stalin acknowledge support from the Indo-Swiss Personnel Exchange Programme INT/SWISS/ISJRP/PEP/P-01/2012. COSMOGRAIL is financially supported by the Swiss National Science Foundation (SNSF). We thank the referee Peter Schneider for his timely comments that helped to improve this paper.
\end{acknowledgements}

\bibliographystyle{aa}
\bibliography{biblio}
\end{document}